\documentstyle[12pt]{ioplppt}          

\newcommand{\be}{\begin{equation}}
\newcommand{\ee}{\end{equation}}
\newcommand{\bea}{\begin{eqnarray}}
\newcommand{\beas}{\begin{eqnarray*}}
\newcommand{\eea}{\end{eqnarray}}
\newcommand{\eeas}{\end{eqnarray*}} 
\newcommand{\ba}{\begin{array}}
\newcommand{\ea}{\end{array}}

\begin{document}

\jl{1}

\title{Factorization, ladder operators and isospectral structures.}
\author{A. P\'erez-Lorenzana.\ftnote{1}{e-mail address: 
abdel@fis.cinvestav.mx}} 
\address{  
Departamento de F\'{\ii}sica, \\ Centro de Investigaci\'on y de Estudios 
Avanzados del IPN, \\ Apartado Postal 14-740, 07000, M\'exico D. F.,
M\'exico.}

\begin{abstract}
Using the modified factorization method employed by Mielnik for the
harmonic oscillator, we show that isospectral structures associated with
a second order operator $H$, can always be constructed whenever $H$ could
be factored, or exist ladder operators for its eigenfunctions. Three
examples are shown, and  properties like completeness and integrability
are discused for the general case. 
\end{abstract}


\section{Introduction.} 
 
In the context of one or tree-dimensional Quantum Mechanics, the
principal problem is to solve the eigenvalue problems associated with the
Schr\"odinger equation when the particle is under the influence of a
general potential. The two most commonly used exact methods to determinate
the eigenfunctions and the eigenvalues (once we have separated the
differential equation) are the method of orthogonal functions and the
factorization method. Nevertheless, the number of potentials which have
exact solution by these ways is quite small.

In recent years, Abraham and Moses \cite{aym} showed one algorithm that
allows us to  generate  exactly solvable potentials from another 
potential which has exact solutions, both with the same eigenvalues, but
adding or subtracting a  finite number of them through the systematic use
of the Gelfand-Levitan equation \cite{gyl}.

On the other hand, Mielnik \cite{mielnik} observed that for the harmonic
oscillator  the same isospectral class of potentials that Abraham and
Moses obtained, may be easily built using the ladder operators. On the
same way, Nieto \cite{nieto} showed that into  supersymmetric Quantum
Mechanics, the natural factorization  that of the bosonic or fermionic
one-dimensional Hamiltonian could produce isospectral Hamiltonians too.
Nevertheless, the use of the factorization scheme to produce new classes
of solvable potentials from other one is not restricted to the two kinds
before, as we will show.

The present work has been divided into the following sections, in section
2, we resume the procedure used by Mielnik in order to find an isospectral
family to the harmonic oscillator. Along section 3, we show how this
property is not exclusive of that case, but all the second order
differential  operators which may be factored by ladder operators have it.
Moreover, this property exists always that a second order operator can be
factored in a more general way, not necessarily by ladder operators or in
a kind of factorization like the supersymmetry one. Section 4 contains
three different examples of this techniques, the free particle in one and
three dimensions and the isotropic oscillator. Finally, our conclusions
are summarized in section 5.

\section{The oscillator's isospectral family}

The first family of isospectral operators built  in a different way from
that of Abraham and Moses \cite{aym}, was obtained for the one-dimensional 
harmonic oscillator, using a variant of the factorization method proposed
by Mielnik \cite{mielnik}. In this case, the Hamiltonian
\be
H = -{1\over 2}{\d^2 \over {\d x^2}}+{1\over 2}x^2 
\ee
whose eigenvalues are $E_n =
n+\frac{1}{2}$, is factored by the creation and annihilation operators
\be
a = \frac{1}{\sqrt{2}} \left(\frac{\d}{\d x} + x \right), \qquad 
a^\dagger = \frac{1}{\sqrt{2}} \left(-\frac{\d}{\d x} + x \right), 
\ee
in such a way that $H= a^\dagger a + \frac{1}{2} = a a^\dagger -
\frac{1}{2}$.  The method considers the more general operators
\be 
b = \frac{1}{\sqrt{2}} \left(\frac{\d}{\d x} + \beta(x) \right), \qquad 
b^\dagger = \frac{1}{\sqrt{2}} \left(-\frac{\d}{\d x} + \beta(x) \right), 
\ee
which must be such that $ H = b b^\dagger - \frac{1}{2} $.

Moreover $\beta(x)$ satisfy the Ricatti equation $\beta ' + \beta^2 = x^2
+ 1$, which have the particular solution $\beta_0 = x$. Then the general
solution can be obtained putting $\beta = x + \phi(x)$. This yields
\be
\phi(x) = { \e^{-x^2} \over{ \lambda + \int^x_0 \e^{-y^2}\, \d y }}.
\ee
where $\lambda$ is a parameter. 
Now, we write down the inverse product $b^\dagger b$
\be
H_\lambda \equiv b^\dagger b +{1\over 2} = -{1 \over 2} + 
{1 \over 2} x^2 - \phi'(x) = H - \phi'.
\ee
Hence, $H_\lambda$ is a certain new family of Hamiltonians. Moreover, if 
we define  $\psi_n = b^\dagger \varphi_{n-1}$, where $\varphi_n$ are the
eigenfunctions of $H$, thus, $\psi_n$ are the eigenfunctions of
$H_\lambda$, with eigenvalues $E_n = n + \frac{1}{2}$
\be
H_\lambda \psi_n = b^\dagger\left( H + 1\right) \varphi_{n-1} = (n+
\frac{1}{2}) \psi_n.
\ee
Therefore, $H$ and $H_\lambda$ are isospectral, except for one missing
element, the ground state of $H_\lambda$, which should be orthogonal to
all of $\psi_n$
\be
\left( \psi_0, \psi_n\right) = 0 \Longrightarrow b \psi_0 = 0.
\ee
We will call to $\{ H, \varphi_n \}$ and $\{ H_\lambda, \psi_n\}$ {\it
isospectral structures} (IS). Some other  usual properties of the
harmonic oscillator are inherited for its IS, like the coherent states, as
has been shown before \cite{rosas}. As it can be seen immediately, this is
not the only case for which there must exist IS \cite{fer}. The existence
of IS for a second order differential operator $H$ is  related with
the existence of a factorization scheme, as we shall show below.

\section{IS by the Factorization Scheme.}\label{sec3}
\setcounter{equation}{0}
 
We begin by establishing the immediate generalization of the Mielnik's
results. Consider the second order differential  operator 
\be
H = P(x) {\d^2\over{\d x^2}} + Q(x){\d\over{\d x}} + R(x) \label{3.1}
\ee
and its eigenvalue equation
\be
H\varphi_n = \lambda_n \varphi_n. \label{3.2}
\ee
By simplicity, we will supose that $n$ is an discrete index, and that
$\lambda_n$ has a minimum value. We assume that there exist the first 
order ladder operators, then we may write them as
\be
A^+_n  =  \alpha_n(x) + \beta_n(x) {\d\over{\d x}}, \qquad
A^-_n  =  \gamma_n(x) - \delta_n(x) {\d\over{\d x}}, \label{3.3}
\ee
where
$A_n^\pm \varphi_n = c^\pm_n \varphi_{n\pm 1}$, with $c^\pm_n$ constants,
and they are such a way that factored to $H$ into
\bea
H & = & A_{n-1}^+ A_n^- + k_n^1\label{3.4.a}\\
 & = & A_{n+1}^- A_n^+ + k_n^2, \label{3.4.b}
\eea
the constants $k_n^1$ and $k_n^2$ are related with $\lambda_n$ by
$\Delta\lambda_n \equiv \lambda_{n+1} - \lambda_n = k_{n+1}^1 -
k_n^2.$ 
Notice that of equations (\ref{3.3}), (\ref{3.4.a}) and (\ref{3.4.b})
 $\alpha_n,\ \beta_n$, $\gamma_n $ and $\delta_n$ satisfy
\bea
-\beta_{n-1}\delta_n & = & P ,\nonumber \\ \beta_{n-1} \gamma_n -
\beta_{n-1} \delta_n' - \alpha_{n-1} \delta_n & = & Q, \label{3.6}\\
\alpha_{n-1} \gamma_n + \beta_{n-1} \gamma_n' + k_n^1 & = & R,
\nonumber \\ \alpha_{n-1} \gamma_n - \delta_{n} \alpha_{n-1}' +
k_n^2 \qquad & = & R. \nonumber
\eea
Next, consider the more general operators
\be
a^{(+)}_n  = \tilde \alpha_n(x) + \beta_n {\d\over{\d x}}, \qquad 
a^{(-)}_n  = \tilde \gamma_n(x) - \delta_n {\d\over{\d x}}. \label{3.8}
\ee
obtained by the changes $\alpha_n \rightarrow  \tilde\alpha_n\equiv
\alpha_n +  \eta_n(x)$ and  $\gamma_n \rightarrow  \tilde\gamma_n\equiv
\gamma_n +  \nu_n(x)$ in (\ref{3.3}), with $\eta_n$ and $\nu_n$ functions.
Then, we will have the two following cases

{\bf Case I. }
Assuming that $H$ is factored by $a^{(\pm)}_n$ into the form
\be
H = a_{n-1}^{(+)} a_n^{(-)} + k^1_n, \label{3.9}
\ee
and using the equations (\ref{3.8}) and (\ref{3.6}) we obtain for $\eta_n$
and $\nu_n$
\be
\eta_{n-1} = \frac{\beta_{n-1}}{\delta_n} \nu_n, \label{3.10}
\ee
and the Ricatti equation for $\nu_n$
\be
\beta_{n-1} \nu_n' + \left( \frac{\beta_{n-1}}{\delta_n} \right) \nu^2_n +
\left[ \alpha_{n-1} + \gamma_n \left( \frac{\beta_{n-1}}{\delta_n}
\right)\right]  \nu_n = 0, \label{3.11}
\ee
whose general solution is
\be
\nu_n(x) = \e^{-{\cal I}_n} \left[ \lambda + \int \! \d x\,
\frac{\e^{-{\cal I}_n}}{\delta_n} \right]^{-1}; \quad  {\cal  I}_n \equiv
\int \! \d x \left[ \frac{\alpha_{n-1}}{\beta_{n-1}} +
\frac{\gamma_n}{\delta_n} \right]. \label{3.12}
 \ee
Hence, the factorization of $H$ into the form (\ref{3.9}) is not unique.
Moreover, building the inverted product $a_{n+1}^{(-)} a_n^{(+)}$, we may
find  the new one--parametric family of operators
\be 
H_n^\lambda\equiv a_{n+1}^{(-)} a_n^{(+)} = H - \left\{
\delta_{n+1}\eta_n' + \beta_n \nu_{n+1}'\right\}  , \label{3.14}
\ee
which as we can immediately see, has identical spectra to that of $H$, 
even though their eigenvectors are different. If we define
\[ \psi_n = a_{n+1}^{(-)} \varphi_{n+1}\]
then
\be
 H^\lambda_n \psi_n = a^{(-)}_{n+1} \left(H - \Delta \lambda_n\right)
\varphi_{n+1} = \lambda_n\psi_n. \label{3.16}
\ee
Therefore, $\psi_n$ are the eigenvectors of $H_n^\lambda$ and one state
could be added to the spectra $\lambda_n$.

Note that in the more general case $H^\lambda_n$ defines a different
operator  for different $n$. Hence, to be precise, $H$ and $H_n^\lambda$
share only one element of their spectra: $\lambda_n$. In fact, the
functions $\{\psi_n\}$ are not orthogonal in general, because equation
(\ref{3.16}) gives
\[ 
\left( \lambda_n - \lambda_m\right) \left(\psi_m,\psi_n\right) =
\left(\psi_m, \Delta_{mn}\, \psi_n\right), 
\]
with $\Delta_{mn} = H^\lambda_n - H^\lambda_m$, a non vanishing scalar
function. However, when the index $n$ is  different to that  defined by
equation (\ref{3.2}), like the index $l$ for the angular momentum into the
Schr\"odinger radial equation with central potentials, the dependence of
$H^\lambda_n$ in $n$ may be seen as a result of the action of one potential,
which is coupled in a different way for each state, like the effective
central potential, with the orthogonality and completeness established by
the Sturm--Liouville theory \cite{arfken}. 

On the other hand, when ${a^{(-)\dagger}_n}= a_{n-1}^{(+)}$, then
$\{\psi_n\}$ will have the integrability properties of $\{\varphi_n\},$
except for a change  on the normalization
\[\left(\psi_n,\psi_n\right)= \left(\lambda_n -
k^1_n\right)\left(\varphi_n,\varphi_n\right). \]
It is clear that if the ladder operators do not depend on the index $n$,
all the properties of $\{\varphi_n\}$ must be inherited by $\{\psi_n\}$,
and  probably there is another  eigenstate of $H^\lambda$ obtained as the
solution to
\[ (\chi,\psi_n)=0 \quad \rightarrow \quad a^{(+)}\chi = 0. \]

{\bf Case II. }
The above results may be obtained again, when $a^{(\pm)}_n$ factored $H$
into
\be
H= a^{(-)}_{n+1}\, a^{(+)}_n + k^2_n.
\ee
In such a case, we find
$\nu_{n+1} \beta_n = \delta_{n+1}\eta_n $
and the Ricatti equation for $\eta_n$ 
\[\left[ \gamma_{n+1} + \left( \frac{\delta_{n+1}}{\beta_n}\right)
 \alpha_n \right] 
\eta_n + \left(\frac{\delta_{n+1}}{\beta_n}\right)  \eta^2_n -
\delta_{n+1}\eta'_n = 0, \] whose general solution is non trivial, as in
the case I. Hence, we may define the new family of operators, like
(\ref{3.14})
\be
\tilde H^\lambda_n \equiv a^{(+)}_{n-1} a^{(-)}_n + k^1_n = H  + \left\{
\beta_{n-1}\nu'_n + \delta_n\eta_{n-1}\right\}, \label{3.21}
\ee
whose eigenfunctions 
\be
\Phi_{n} \equiv a_{n-1}^{(+)} \varphi_{n-1}
\ee
have the same eigenvalues that of $\varphi_n$. But now, the spectra has
not  the ground state.  Note that the comments for the case I  still hold
in the present one.

Once we know the existence of IS for $H$, we must emphasize that the
constriction  on $A^\pm_n$ to be ladder operators and factored to $H$ at
the same time, is not entirely necessary, IS can be built if just one of
them  holds as we will show below.

Let $A^\pm_n$ be ladder operators, in the more general case they do not
factorize to $H$ \cite{apl}, instead, 
\be
D_n \equiv  A^-_{n+1} A^+_n  \qquad \mbox{and}\qquad
E_n \equiv  A^+_{n-1} A^-_n  \label{4.1}
\ee
define two different operators, which obviously share eigenfunctions with
$H$, but not eigenvalues. Even now, the algorithm discussed above can be
reproduced to build IS associated with  $\{ D_n,\varphi_n\}$ and
$\{E_n,\varphi_n\}$, but it is important to note that we do not have IS
for $H$ in a natural way.
Nevertheless, when $A_n^\pm$ and $H$ are related by
\be 
\left ( \ba{l} D_n \\ E_n\ea \right) = f(x)\left( H - \lambda_n\right) +
\left(\ba{l} d_n\\ e_n\ea\right), 
\ee
$d_n,e_n$ being constants, it is an easy task to find IS for $H$, as we
shall note in the examples below.

Now, if we assume that $H$ is factorizable, not necessarily by ladder
operators, then there are two first order operators $B_n$ and $C_n$ which 
satisfy
\be H = B_n C_n + k_n\ee
where $k_n$ is a constant. In this case, it is always possible to follow
the usual procedure in order to find IS for H. Moreover, as $F_{(n)}\equiv
C_n B_n$ is in general a different operator to $H$, IS may be built from
it in a direct way, whenever  $F_{(n)} C_n = C_n (B_n C_n) = C_n
H_{(n)}$. 
Hence, $ \xi_n \equiv C_n \varphi_n  $  are eigenfunctions of $F_n$ with
eigenvalue $\lambda_n$. Thus, we have established that {\it is sufficient
with factored one second order operator $H$, to assure that one IS
$\{H^\lambda,\xi_n\}$ associated with $\{H,\varphi_n\}$ must  exist},
even when the spectra is continuos.

In the case when $H$ does not depend in $n$, neither $B_n$, $C_n$ and
$k_n$ do, and then the family $H^\lambda$ will be defined in a proper way.
Moreover, in such a case, the completeness and orthogonality of
$\{\xi_n\}$ must  be insured by the properties of the eigenfunctions of
$H$, or by the Sturm--Liouville theory.

Finally, the present approximation is not the more general. It is possible
to generalize the four functions $\alpha_n$, $\beta_n$, $\gamma_n$ and
$\delta_n$ at the same time, but this scheme modifies $H$ in something more
than one scalar function, in such a way that a new family of isospectral
operators  to $H$ can be found, but the new operators might have not a
direct physical interpretation, and just have mathematical sense. These kind
of IS  will not be considered here.

\section{Examples.}
\setcounter{equation}{0}

We shall illustrate the above techniques for several examples: the free
particle in one and three dimensions and the isotropic oscillator.

\subsection{The Free Particle in one dimension.}
    
We first consider the most simple example: a free particle, that is, one
for which $V(x)=0$. The Schr\"odinger equation is
\be
H\varphi_k = -{\d^2\over \d x^2} \varphi_k = k^2 \varphi_k
\ee
where $k^2 = 2mE/\hbar^2$. Obiously, $\varphi_k$ are $\sin (kx)$ or $\cos
(kx)$. $H$ may be factored into $H = A A^\dagger = A^\dagger A $, with 
$A = {\d\over \d x}$.

Next, consider the generalized operator
\be a = {\d\over \d x} + \nu (x). \ee
Thus, if $H = a a^\dagger$ this leads to
\be \nu (x) = {1\over \lambda + x}. \ee
Hence, we may propose the new operator
\be H^\lambda = a^\dagger a = -{\d^2\over \d x^2} + V_\lambda (x),\ee
where
\be V_\lambda (x) \equiv 2\nu'(x) = {-2\over (\lambda + x) ^2}. \ee
The above potential has one singularity in $x_0 = -\lambda$ and vanishes
for $x\rightarrow \pm \infty$. The eigenfunctions of $H^\lambda$ are
\be
 \chi_k\equiv a^\dagger \varphi_k \Rightarrow H^\lambda \chi_k = k^2
\chi_k. 
\ee
These eigenfunctions are orthogonal
\be (\chi_k,\chi_{k'}) = (a a^\dagger \varphi_k, \varphi_{k'}) =
k^2(\varphi_k,\varphi_{k'}). \ee
Moreover, as $a\chi = 0$ has the only solution $\chi = \nu (x)$ and
$\chi_0(x) = \nu (x)$, then, $\{ \chi_k \}$ is a complete set, and like
$\{\varphi_k\}$, they are an external basis for the Hilbert space. Other
IS can be found if we assume now that $H = a^\dagger a$, but it is
essentially of the same kind, because in this case we will obtain $\nu =
(\lambda - x)^{-1}$.

Unlike  the algorithm of Abraham and Moses \cite{aym}, when we consider the
particle in a box, the boundary conditions which satisfy the wave
functions $\varphi$, are not  inherited by the eigenvectors of the new
Hamiltonians. Therefore, the spectra of one particle in a box  with
potential $V_\lambda(x)$, must be found  by making that the general
solution satisfies the boundary conditions  of the box, which give us
different quantization conditions on $k$ to that of the box without
potential, and then the new spectra will not be  equal to the particle in
a box.

The present example may be extended straightforward to the three
dimensional problem. However it is more illustrative the use of the ladder
operators along one of the indices which does not denote the energy level.

\subsection{The Free Particle in three dimensions.}
 
The Schr\"odinger equation for the free particle in three dimensions
takes the form
\be 
H\psi = -{\hbar^2\over{2 m}} \nabla^2 \psi = E\psi. \label{6.1}
\ee
Equation (\ref{6.1}) can be separated in spherical coordinates,
introducing  $\psi (r,\theta,\phi) = R_{l}(kr) Y_{lm}(\theta,\phi)$. The
radial equation becomes the spherical Bessel equation \be {1\over
{\rho^2}} {\d\over{\d \rho}}\left(\rho^2{\d R_l(\rho)\over \d\rho}\right) 
+ \left(1 - {l(l+1)\over \rho^2}\right) R_l(\rho) = 0,\label{6.3}
\ee
where $k^2 = 2mE/\hbar^2$, and $\rho= kr$, it has as general solution one
linear combination from the spherical Bessel functions $j_l(\rho)$ and
$n_l(\rho)$ \cite{schiff}.
 Although the radial equation defines the energy, it  may be
written  down as an eigenvalue equation for $l$, hence, the procedure
showed in the section \ref{sec3} can be applied here. Consider the
recurrence relations \cite{arfken} 
\be 
f_{l-1} + f_{l+1}  = {2l + 1\over x}f_l \quad\mbox{ and} 
\quad l f_{l-1} - (l+1) f_{l+1} =  (2l
+ 1)f'_l \label{6.5} \ee
satisfied by $j_l$ and $n_l$. Using the last equation we may build the
ladder operators
\be 
 A^+_l\equiv {l\over\rho} - {\d\over \d\rho} \quad\mbox{ and} \quad
A^-_l\equiv {l+1\over\rho} + {\d\over \d\rho} \quad
\mbox{ such that} \quad A^\pm_l f_l = f_{l\pm1} \label{6.6}
\ee
which factored $H_l$ into $H_l = A^+_{l-1}A^-_l = A^-_{l+1}A^+_l$. Hence,
the IS associated with $H_l$ will be

{\bf I.-}
 The new family of hamiltonians 
$H^{\lambda_1}_l \equiv H_l + V^{\lambda_1}_l(\rho)$,
with
\be
V_l^{\lambda_1}(\rho) = 2 {\d\over \d\rho} \nu_{l+1}^{\lambda_1}(\rho)
\quad \mbox{and}\quad \nu_l^{\lambda_1}(\rho) ={\rho^{2l}}
\left[{\lambda_1 - {\rho^{2l+1}\over({2l+1})}} \right]^{-1},
\ee
whose eigenfunctions are
\be
\psi^{\lambda_1}_{l-1}(\rho) = \left[{\d\over \d\rho} + {l+1\over \rho} +
\nu_l^{\lambda_1}\right] R_l (\rho).
\ee
As can be seen, if $\lambda_1<0$, then neither $\nu_l^{\lambda_1}$ and
$V_l^{\lambda_1}$ have singularities, and therefore neither
$\psi^{\lambda_1}_{l}$ has. Moreover, as  $\nu_l^{\lambda_1}\rightarrow 0$
when $r\rightarrow \infty$, it  assures the integrability of the
eigenfunctions obtained from $j_l$.

{\bf II.-}
The other one parametric family is 
$H^{\lambda_2}_l \equiv H_l - V^{\lambda_2}_l(\rho)$, where
\be 
V_l^{\lambda_2}(\rho) = 2 {\d\over \d\rho} \nu_l^{\lambda_2}(\rho) \quad
\mbox{and}\quad \nu_l^{\lambda_2}(\rho) =\rho^{-2l} \left[{{\lambda_2 +
{\rho^{-2l+1}\over{(-2l+1)}}} }\right]^{-1},
\ee
and in this case the eigenfunctions are
\be
\psi^{\lambda_2}_{l+1}(\rho) = \left[-{\d\over \d\rho} + {l\over \rho} +
\nu_{l+1}^{\lambda_2}\right] R_l (\rho).
\ee
Now, as $\nu_0^{\lambda_2}=(\rho+\lambda_2)^{-1}$, it has a singularity in
$\rho =-\lambda_2$ while $\nu_l^{\lambda_2}$ has one in $\rho=0$ if
$\lambda_2<0$. For $\rho<< 1$ the limiting values for $\nu_l^{\lambda_2}$
have the form
\be
 \nu_l^{\lambda_2}\sim -{2l-1\over\rho}, 
\ee
and therefore the respective limiting form of $\psi^{\lambda_2}_{l+1}$ is 
\be 
\psi^{\lambda_2}_{l+1}\sim \left[-{\d\over \d\rho} + {l\over \rho}
-{2l+1\over\rho} \right] R_l =  -A_l^- R_l = -R_{l-1},
\ee
moreover, considering the limiting values of $j_l$ \cite{arfken}
\be 
j_l(\rho)\sim {\rho^l\over(2l+1)!!}
\ee
we note that $\psi^{\lambda_2}_{l}$ has not singularities for $l\neq 0$
and $\lambda_2<0$, while $\psi^{\lambda_2}_{1}$ has one in $\rho =0$ like
$1/\rho$, but it do not appear if we use $R_0= n_0$ instead $j_0$. On the
other hand, as $\nu^\lambda_l$ vanishes in the limit  $r\rightarrow
\infty$, then the eigenvectors always go to zero at infinity.

\subsection{The Isotropic Oscillator.}
 
The isotropic three-dimensional harmonic oscillator has the characteristic
that  when the Sch\"odinger equation is  separated in spherical
coordinates, the eigenvalues of the energy are defined in function of two
indices, the angular momentum $l$ and that of the radial equation, in such
a way that  IS may be build by ladder operators associated to each index.
The radial equation for this problem takes the form \cite{schiff}
\be 
H_lu_{nl} = \left[-{\d^2\over \d r^2} + r^2 + {l(l+1)\over r^2}\right]
u_{nl}(r) = \varepsilon_{nl}u_{nl} \label{7.1}
\ee
with $u_{nl}(r) = rR_{nl}(r)$, $\varepsilon_{nl} = (4n + 2 l + 3)$.
 The solution to (\ref{7.1}) is
\be  
u_{nl}(\rho) = \rho^{(l+1)/ 2} e^{-{\rho/ 2}} L_n^{l+{1\over2}}(\rho)
 \label{unl}
 \ee
where $ L_n^{l+{1/ 2}}(\rho)$ are the associated Laguerre polynomials, and
$\rho= r^2$. From the recurrence relations \cite{arfken,gradshteyn}
\beas 
(n+1)L^k_{n+1} &=& (2n + k + 1 -\rho)L^k_n - (n+k) L^k_{n-1} \\
\rho {\d\over \d\rho} L^k_n &=& nL_n^k - (n+k) L_{n-1}^k\\
L_n^{k-1} &=& L_n^k - L_{n-1}^k\\
\rho L_n^{k+1} &=& (n+k) L_{n-1}^k - (n-\rho)L_n^k 
\eeas
we may find the ladder operators over $n$
\be A_n^- = n - \rho{\d\over \d\rho}, \qquad A_n^+ = (n+k+1-\rho) + 
\rho{\d\over \d\rho} \ee
and over $k$
\be 
B_k^- =  k+ \rho{\d\over \d\rho} \qquad B^+ \equiv B_k^+ = 1 - 
\rho{\d\over \d\rho}
\ee
for the $L_n^k$ functions, which were obtained by J. Morales {\it et
al.} \cite{lopezbonilla} in a different way.
 Using the last operators is an easy task to  build
the ladder operators over $n$  or $l$ for $u_{nl}$. Note that we can write
\be
 u_{nl}(\rho) = f_l(\rho) L_n^{l+{1\over 2}}; \qquad 
f_l = \rho^{(l+1)/ 2} \e^{-{\rho/ 2}}.
\ee
Thus, the ladder operators on $l$ become
$a_l^\pm = f_{l\pm 1}\, B^\pm_{l+{1/2}}\, f^{-1}_l$,
which in function of $r$, may be written down as
\be 
a_l^- = \left[ {l\over r} + r + {\d\over \d r} \right], \qquad a_l^+ =
\left[ {{l+1}\over r} + r - {\d\over\d r}\right]. 
\ee

Hence, $(a^-_l)^\dagger = a^+_{l-1}$, and then the integrability of the
new eigenvectors is assured from the integrability of $\{u_{nl}\}$.
Moreover, $H_l$ is factored into $H_l = a^+_{l-1}a^-_l - (2l -1) =
a^-_{l+1} a^+_l - (2l + 3)$. As above, there are two class of IS
associated with $H_l$

{\bf I.-} The family of Hamiltonians
$H^{\lambda_1}_l = H_l + V^{\lambda_1}_l(r)$,
where
\be 
V^{\lambda_1}_{l} = 2{\d\over \d r}\nu^{\lambda_1}_{l+1}; \qquad 
\nu^{\lambda_1}_l = {\e^{r^2} r^{2l}\over \left[\lambda_1 - \int^r_0 x^{2l}
\e^{x^2} \d x\right]}; 
\ee
and  the eigenfunctions of $H^{\lambda_1}_l$ 
\be 
\psi^{\lambda_1}_{nl-1} = \left[{\d\over \d r} + r + {l\over r} +
 \nu^{\lambda_1}_l \right] u_{nl}, 
 \ee
with eigenvalues given by $\varepsilon_{nl}$. It is important to note that
if $\lambda_1\leq 0$ then $V^{\lambda_1}_l \rightarrow 0$ when
$r\rightarrow 0,\infty$; and it has not singularities. After some algebra,
it can be show that the new eigenfunctions  have not singularities, and
they go to zero when $r\rightarrow \infty$.

{\bf II.-} The other isospectral class is formed by the Hamiltonians
$ H^{\lambda_2}_l = H_l + V^{\lambda_2}_l(r)$,
and its eigenfunctions
\be 
\psi^{\lambda_2}_{nl+1} = \left[-{\d\over \d r} + r + {l+1\over r} +
 \nu^{\lambda_2}_{l+1} \right] u_{nl}, 
 \ee
where now, 
\be V^{\lambda_2}_l = -2{\d\over \d r}\nu^{\lambda_2}_l; \qquad 
\nu^{\lambda_2}_l = {\e^{-r^2} r^{-2l}\over \left[\lambda_2 - 
\int^\infty_r x^{-2l}
\e^{-x^2} \d x\right]}. 
\ee
In this case, if $\lambda_2<0$ and $l\neq 0$, then $\nu^{\lambda_2}_l$ and
$V^{\lambda_2}_l \rightarrow 0$ when $r\rightarrow \infty$, but
$\nu^{\lambda_2}_l$ has a singularity in $r=0$ like $1/r$ while
$V^{\lambda_2}_l$ diverges like $1/r^2$. Again, when we analize
$\nu^{\lambda_2}_l$ around $r=0$ we find
\be 
\nu^{\lambda_2}_l \sim -2 \left[ r + {l \over r} \right]
\ee
and then
\be
\psi^{\lambda_2}_{nl+1}\sim -\left(u_{nl-1} + {u_{nl}\over r}\right)
\ee
but because (\ref{unl}) the eigenfunctions have not divergences. For $l=0$
one can easily check that if $\lambda_2<0$ or $\lambda_2>\sqrt{\pi}/2$, 
$\nu^{\lambda_2}_0$ has not singularities and so neither $V^{\lambda_2}_0$
and $\psi^{\lambda_2}_{n0}$ have.

It is important to note, that both cases above are very different
to that reported by Fern\'andez \cite{fertes} and Dongpei \cite{dongpei}.

An analogous procedure may be followed to build the ladder operators on
$n$ for $u_{nl}$, though we should note that when  $n\rightarrow n+1$, the
energy level is moved two levels, therefore, we must consider two kinds of
IS, depending on what part of the spectra we are moving. In order to
simplify, we shall consider only the case $l=0$. So, the ladder operators
on $n$ may be written as
\be 
a_n^- = \left[n + {1\over 2} - {r^2\over 2} - {r\over 2}{\d\over
\d r}\right],
\qquad
a_n^+ = \left[n + 1 - {r^2\over 2} + {r\over 2}{\d\over \d r}\right]; 
\ee
then $(a_n^-)^\dagger = a_{n-1}^+$, it assures the integrability of the
eigenvectors associated with the new structures. Note that in this case,
$H$ is not factored by $a_n^\pm$, instead we have
\be 
E_n\equiv a^-_{n+1}a^+_n = {r^2\over 4}\Big[ H_0 - \varepsilon_{n0} \Big]
+ \big(n+1\big)\Big(n+{3\over 2}\Big), 
\ee
\be 
D_n\equiv a^+_{n-1}a^-_n = {r^2\over 4}\Big[ H_0 - \varepsilon_{n0} \Big]
+ n \Big(n+{1\over 2}\Big). \label{8.3}
\ee
So, we should build IS for $D_n$ and $E_n$ and from them for $H_0$. 
Following the algebra, we find, for example for $E_n$
\be 
E_n^\lambda = D_{n+1} + V^\lambda_n \ , \qquad V^\lambda_n = r{\d\over  
\d r} \nu^\lambda_n ,\label{8.4}
\ee
where
\be 
\nu^\lambda_n = {\e^{-r^2}\, r^{4n+5}\over \left[\lambda - 
2 \int^r_0 \e^{-x^2} x^{4(n+1)}\d x\right] }.
\ee
Hence, if $\lambda<0$, then  $V^\lambda_n$, $\nu^\lambda_n$  and the
respective eigenfunctions corresponding to the eigenvalue $e_n =
(n+1)(n+3/2)$, which become
\be 
\psi_n = \left[n+1 -{r^2\over 2} + \nu^\lambda_n + 
{r\over 2}{\d\over \d r}\right] u_{n0},
\ee
have not singularities and go to zero at infinity.
Moreover, IS for $H$ may be found from (\ref{8.4}) using (\ref{8.3}) to
give
\be 
E^\lambda_n = {r^2\over 4}\Big[ H- \varepsilon_{n+1,0}\Big] + e_n +
V^\lambda_n, 
\ee
and writting the last equation as
\be
H^\lambda_{n+1}\chi_{n+1} \equiv  \left[ H+{r^2\over 4} V^\lambda_n
\right] \chi_{n+1} = \varepsilon_{n+1,0}\, \chi_{n+1}, 
\ee
where $\chi_{n+1} = \psi_{n}$,  we note that $H^\lambda_n$ is a family of
semi-isospectral hamiltonians to $H$.

\section{Concluding Remarks}
 
As we have shown, the existence of isospectral families of potentials is 
associated with the general factorization of the Hamiltonian $H$, even in
the case  when the operators which factorize to $H$ are not ladder
operators, and regardless $H$ may has a continuos spectra $\{\lambda_n\}$.
Also, there are IS when exist ladder operators, even though $H$ is not
explicitly factorized by them, but in this case, as we have shown above,
they are not associated with $H$ in a natural way.

 In the most general case, the eigenvectors $\{\phi_n\}$ of the new
 hamiltonians $H^\lambda_n$  do not  inherit the property of satisfying
 the boundary conditions that the original eigenfunctions $\{\varphi_n\}$
 satisfy, just as happened in the case of the free particle in one
 dimension, and when $a_n^{(-)\dagger} = a_{n-1}^{(+)}$, $\{\phi_n\}$ will
 have the integrability properties of $\{\varphi_n\}$. Moreover, the way
 to determinate the IS, is not unique, as in the case of the isotropic
 harmonic oscillator. In general, it is possible that exist more than one
 kind of IS associated to one operator, depending on how many independent
 ways to factored it can be found,  and with that, the structures like the
 coherent states for the harmonic oscillator's isospectral family, related
 to the IS should be studied very carefully.

On other hand, the algorithm discussed here has an strictly mathematical
sense. Sometimes, as it happen in the case of the free particle, the new
eigenvectors could  have singularities and have not physical interest, but
still this kind of functions could be important if we have bounded
potentials, therefore they should not be rejected.

\ack

The author acknowledges CONACyT (M\'exico) for financial support. Thanks
are due to B. Mielnik, D. J. Fern\'andez, H. H. Garc\'{\ii}a, M.
Montesinos--Vel\'asquez and J. L. G\'omez for their helpful discussions and
comments.

\end{document}